\documentclass[preprint]{aastex}

\usepackage{graphicx}
\usepackage{natbib}
\usepackage{psfrag}
\usepackage{amsmath}
\usepackage{epsfig}
\shorttitle{Acetic Acid in Hot Molecular Cores}
\shortauthors{Shiao et al.}

\begin{document}

\title{FIRST ACETIC ACID SURVEY WITH CARMA IN HOT MOLECULAR CORES}

\author{Y.-S. JERRY SHIAO\altaffilmark{1,2}, LESLIE W. LOONEY\altaffilmark{1},
        ANTHONY J. REMIJAN\altaffilmark{3,4}, LEWIS E. SNYDER\altaffilmark{1}, 
	and DOUGLAS N. FRIEDEL\altaffilmark{1}}

\altaffiltext{1}{Department of Astronomy, University of Illinois at
    Urbana-Champaign, Urbana, IL 61801, USA}
\altaffiltext{2}{Physics Department, National Taiwan
University, Taipei 10617, Taiwan; jshiao@phys.ntu.edu.tw}
\altaffiltext{3}{National Radio Astronomy Observatory, 
Charlottesville, VA 2290, USA; aremijan@nrao.edu}
\altaffiltext{4}{Center for Chemistry of the Universe, Department of
    Chemistry, University of Virginia, McCormack Rd., P.O. Box 400319,
    Charlottesville, VA 22904-4319}

\begin{abstract}

Acetic acid (CH$_3$COOH) has been detected mainly in hot molecular cores where
the distribution between oxygen (O) and nitrogen (N) containing molecular
species is co-spatial within the telescope beam. Previous work has presumed
that similar cores with co-spatial O and N species may be an indicator for
detecting acetic acid. However, does this presumption hold as higher spatial
resolution observations become available of large O and N-containing molecules?
As the number of detected acetic acid sources is still low, more observations
are needed to support this postulate. In this paper, we report the first acetic
acid survey conducted with the Combined Array for Research in Millimeter-wave
Astronomy (CARMA) at 3 mm wavelengths towards G19.61-0.23, G29.96-0.02 and IRAS
16293-2422. We have successfully detected CH$_3$COOH via two transitions toward
G19.61-0.23 and tentatively confirmed the detection toward IRAS 16293-2422 A.
The determined column density of CH$_3$COOH is 2.0(1.0)$\times$10$^{16}$
cm$^{-2}$ and the abundance ratio of CH$_3$COOH to methyl formate (HCOOCH$_3$)
is 2.2(0.1)$\times$10$^{-1}$ toward G19.61-0.23. Toward IRAS 16293 A, the
determined column density of CH$_3$COOH is $\sim$1.6$\times$10$^{15}$ cm$^{-2}$
and the abundance ratio of CH$_3$COOH to methyl formate (HCOOCH$_3$) is
$\sim$1.0$\times$10$^{-1}$ both of which are consistent with abundance ratios
determined toward other hot cores. Finally, we model all known line emission in
our passband to determine physical conditions in the regions and introduce a new
metric to better reveal weak spectral features that are blended with stronger
lines or that may be near the 1-2$\sigma$ detection limit.

\end{abstract}


\section{INTRODUCTION}

With high resolution observations, Sgr B2 has been resolved into many regions
of  compact molecular/continuum emission \citep{gaume95}; one of these regions, 
Sgr B2(N-LMH) (i.e., the ``Large Molecule Heimat'') \citep{snyder94,miao95}, is
a  molecule-rich hot core that has historically been the best place to search 
for large and complex molecules. Interstellar acetic acid (CH$_{3}$COOH) was
first detected in the hot  molecular core (HMC) of Sgr B2(N-LMH) with the
Berkeley-Illinois-Maryland Association (BIMA) and Owens Valley Radio Observatory
(OVRO) millimeter-wavelength arrays by \citet{meh97}. The compactness of the 
CH$_{3}$COOH emission in Sgr B2(N-LMH) ($<$3\arcsec) \citep[e.g.,][]{remi02}
highlights the importance of interferometers. To clearly detect the compact
emission features from CH$_{3}$COOH and  other large molecules toward hot cores
embedded in regions of more extended molecular emission, high resolution
interferometric observations are necessary. Thus, following the first
CH$_{3}$COOH detections in Sgr B2(N-LMH) and using the BIMA array,
\citet{remi02} confirmed CH$_{3}$COOH in Sgr B2(N-LMH) and discovered the second
CH$_{3}$COOH source, the hot core toward W51e2.

\citet{remi03} launched an extensive CH$_{3}$COOH survey of 12 galactic HMCs, 
including high-mass ($\geq$ 10 M$_\sun$) and low-mass ($<$ 10 M$_\sun$) sources,
and they found the third CH$_3$COOH source toward the high-mass HMC G34.3+0.15.
Using the IRAM 30-m telescope, \citet{cazaux03} reported the first detection of
CH$_{3}$COOH in IRAS 16293-2422 via the $9_{*,9}-8_{*,8}$ $E$ line at 100.855
GHz. This is the first low-mass star forming region with detected CH$_{3}$COOH
to date, making only a total of four known interstellar sources of CH$_{3}$COOH.
At the conclusion of the search for new sources of CH$_{3}$COOH, it was found
that the three high mass regions containing CH$_{3}$COOH emission share certain
chemical properties. For example, the abundances of CH$_{3}$COOH and its isomer
methyl formate (HCOOCH$_3$) are very different; in Sgr B2(N-LMH), the abundance
of HCOOCH$_3$ is 26 times higher than that of CH$_3$COOH \citep{snyder06}
assuming the emission from CH$_3$COOH and HCOOCH$_3$ are co-spatial within the
telescope beam. In addition, it has been suggested by \citet{remi04} that there
is a relationship between the CH$_3$COOH sources and the spatial separation of
complex oxygen (O) (e.g.\ HCOOCH$_3$ and CH$_3$COCH$_3$) and nitrogen (N)
containing species  (e.g.\ CH$_3$CH$_2$CN and NH$_2$CH$_2$CN). At the limit of
the current observational resolution of these surveys, CH$_3$COOH emission is
only detected where the emission from O and N-containing species is co-spatial.
In contrast, there have not been any CH$_{3}$COOH detections in  HMCs with
reported spatial separation of complex O and N-containing species, e.g. Orion KL
\citep[e.g.][]{sutt95} and W3(OH) \citep[e.g.][]{wyro99}, even though the
observed separation is very dependent on source distance. 

Gas phase and grain surface chemistry may both play critical roles in the
formation of CH$_{3}$COOH and other large molecules in the ISM
\citep{hase92,garr06,garr08}.  However, the exact pathways of CH$_{3}$COOH
formation in gas phase or on grain  surfaces in astronomical environments are
still not clear. The apparent spatial coincidence of the emission of complex O
and N-containing species with the CH$_3$COOH emission may suggest a strong
correlation with CH$_3$COOH formation  \citep[e.g.,][]{remi04}. Another
possibility is that CH$_{3}$COOH  and large N-bearing species share
approximately the same time scale to form in hot cores. Nevertheless, it is
important to investigate the spatial relationship between CH$_{3}$COOH and the
distribution of other complex molecular species. Thus, we also observed methyl
formate (HCOOCH$_3$) and ethyl cyanide (CH$_3$CH$_2$CN) as proxies to determine
the spatial distribution of O and N-bearing species toward our sources in this
survey. 

To search for regions of CH$_{3}$COOH emission and continue to determine its
spatial relationship with  the distribution of other complex molecular species,
we have used the high  sensitivity of the Combined Array for Research in
Millimeter-wave Astronomy (CARMA)  to carry out a CH$_{3}$COOH survey toward
three more hot cores. The observed sources are  G19.61-0.23, G29.96-0.02 and
IRAS 16293-2422 and are described in more detail in $\S$2. A complete
description of the CARMA observations are described in $\S$3; the results are
presented in $\S$4; and in $\S$5 we describe the analysis routine and discuss
the implication of these observations with future surveys and chemical formation
models.

\section{OBSERVED SOURCES}

\subsection{G19.61-0.23 \& G29.96-0.02} 

G19.61-0.23 (hereafter, G19) is a molecule-rich ultracompact \ion{H}{2} region,
which includes a high-mass HMC. Numerous molecules have been detected in this
region, including OH \citep{matt77,gar85}, H$_2$O \citep{genz77}, CH$_3$OH
\citep{kale94,lari99},  NH$_3$ \citep{gar98}, CS \citep{shir03,wu03,lari99}, HCN
\citep{wu03} and CO \citep{hofn00}. G29.96-0.02 (hereafter, G29) is another
high-mass  ultracompact region that also has been observed in many molecular
species including CH$_3$CN, C$^{18}$O, CH$_3$OH,  CH$_3$CH$_2$CN, C$^{34}$S,
CH$_3$OCH$_3$, HCOOCH$_3$, $^{34}$SO, SO$_2$, HC$_3$N, H$_2$CS, C$_2$H$_5$OH and
SiO \citep[see e.g.][and references therein]{cesa98,olmi03,beut07}. One of the
more recent surveys toward both sources was conducted by \citet{font07} with the
Institute de Radioastronomie Millim\'{e}trique 30-m (IRAM) telescope.  These
observations targeted  several emission features between 1 and 3 mm wavelengths
of CH$_3$CH$_2$CN,  vinyl cyanide (CH$_2$CHCN) and dimethyl ether
(CH$_3$OCH$_3$). Specifically, the observations by \citet{font07} were conducted
to further investigate whether the  so called ``chemical differentiation'' seen
toward Orion and W3 is a general  characteristic of high mass HMCs. However, no
``chemical differentiation'' was observed within the spatial limits defined by
the size of the synthesized beam of their observations.

Toward G19, from the observations of a series of CH$_3$CH$_2$CN transitions, 
\citet{font07} determined a rotational temperature of T$_{rot}$=116(12) K, and 
from that a source averaged column density of $N_S$=2.2(0.3)$\times$10$^{16}$
cm$^{-2}$. From a series of CH$_3$OCH$_3$ transitions that would be used to
represent  the excitation and distribution of other O-bearing molecules like
HCOOCH$_3$, \citet{font07} determined a T$_{rot}$=158(17) K, and
$N_S$=2.0(0.2)$\times$10$^{15}$ cm$^{-2}$. Note that the half power beam width
(HPBW) of the IRAM 30-m telescope ranged in these 1 mm to 3 mm observations from
$\sim$12\arcsec to $\sim$22\arcsec, respectively. In contrast, using archival data
from the Submillimeter Array (SMA), \citet{wu09} found a much higher temperature
for the hot core regions. Using methyl cyanide (CH$_3$CN) at a spatial
resolution of 2\farcs6$\times$2\farcs1, they found T$_{rot}$$\sim$552 K and the
beam averaged column density, $N_S$$\sim$3.4$\times$10$^{16}$ cm$^{-2}$ 
without optical depth corrections.

In a dedicated CH$_3$COOH survey of high mass HMCs by \citet{remi04}, HCOOCH$_3$
(methyl formate),  CH$_3$CH$_2$CN (ethyl cyanide),  CH$_3$OH (methanol) and
HCOOH (formic acid) were all detected at a spatial resolution of $\sim$5\arcsec~
toward G19.  There was also an indication of weak CH$_3$COOH emission at 111.507
GHz, at the 1$\sigma$ detection limit of the BIMA array.  Based on the detected
CH$_{3}$COOH sources and the upper limits from the non-detections, the abundance
ratio of CH$_{3}$COOH to HCOOCH$_3$ seems universal in high-mass HMCs. If G19
indeed contains CH$_{3}$COOH, the HCOOCH$_3$ detection indicated that the
CH$_{3}$COOH transitions were just slightly lower than the BIMA sensitivity. To
clearly detect the CH$_{3}$COOH transition lines, we needed a higher sensitivity
by a factor of 4, which is provided by CARMA in this survey.

Toward G29, the observations by \citet{font07} of CH$_3$CH$_2$CN transitions
yielded  a T$_{rot}$=121(17) K, and from that a $N_S$=1.5(0.3)$\times$10$^{16}$
cm$^{-2}$. Similarly,  observations of CH$_3$OCH$_3$ transitions yielded a
T$_{rot}$=141(26) K, and  $N_S$=6.2(1.0)$\times$10$^{15}$ cm$^{-2}$.  Another
recent campaign toward G29 was by \citet{beut07} who mapped the spatial
distribution of several complex  molecules with the SMA at a spatial resolution
of $\sim$0\farcs4$\times$0\farcs3. However, no temperature or column density
determinations were made from the emission features of these complex molecules. 
The only reported temperature of the compact emission regions was made from
observations of CH$_3$OH in its vibrationally excited $v_t$=1 state.  From the
relative intensities of the detected emission features, \citet{beut07} reported
an excitation temperature of T$_{ex}$$\sim$340 K with a lower limit of 220 K. 
In addition, this mapping campaign showed the  distributions of CH$_3$OH, 
CH$_3$CH$_2$CN, CH$_3$OCH$_3$ and HCOOCH$_3$ were co-spatial  suggesting that
G29 may be another source of CH$_3$COOH and that an excitation temperature as
high as 220 K may be used in the determination of the column density of these
complex molecules (see $\S$3).

\subsection{IRAS 16293-2422}

IRAS 16293-2422 is one of the most well-studied and observed low-mass star
forming region \citep[e.g.,][]{looney00}. It consists of two clumps denoted as
component A and B (hereafter I16293A and B). The structure of I16293A is much
more complex than I16293B \citep[e.g.][]{chan05}. I16293A contains two
centimeter sources A1 and A2 \citep{woot89}, and two submillimeter sources Aa
and Ab \citep{chan05}. Moreover, I16293A2 may actually be a bipolar ejection
with two components \citep{loin07}. Despite their low masses, both component A
and B have been suggested to be as molecule-rich as high-mass HMCs
\citep[e.g.][]{schoier02,cazaux03}.

Large O-bearing species, including HCOOCH$_3$ and HCOOH, and N-bearing species,
including CH$_3$CN (methyl cyanide), CH$_2$CHCN (vinyl cyanide) and 
CH$_3$CH$_2$CN, have been detected toward this region
\citep{cazaux03,bott04,kuan04,remi06}. With the
IRAM 30-m telescope, \citet{cazaux03} reported the first
detection of CH$_{3}$COOH in this low mass region via the $9_{*,9}-8_{*,8}$ $E$
line at 100.855 GHz. This makes  IRAS 16293-2422 the only low-mass CH$_{3}$COOH
source to date.  However, \citet{cazaux03} did not detect the counterpart,
$9_{*,9}-8_{*,8}$ $A$ line, and the $9_{*,9}-8_{*,8}$ $E$ line is known to be
blended \citep[e.g.,][]{cazaux03}. Comparing the IRAM single dish of a
28\arcsec~beam to an estimated source size of 5\arcsec~for the CH$_{3}$COOH
emission, beam dilution would severely hinder the detection of other weak
CH$_{3}$COOH lines and may even pick up the extended emission from more
distributed molecular species. To further investigate the CH$_{3}$COOH
distribution toward IRAS 16293-2422, we observed two more CH$_{3}$COOH
transitions in this survey with the high resolution of CARMA, which is better
coupled to the assumed source size and is insensitive to extended emission.

\section{OBSERVATIONS}

The CH$_{3}$COOH survey was carried out with the CARMA $\lambda=$ 3 mm receivers
from Spring 2006 to Spring 2008 in the B and C configurations. The individual
angular resolutions are about 1$\arcsec$ and 2$\arcsec$, respectively.
Table~\ref{tbl-src} summarizes the properties of the observed sources, which
includes the source positions, flux calibrators, gain calibrators, distances,
masses and LSR velocities. The six windows in the CARMA correlator were
configured as two wide-band and four narrow-band windows. Each wide-band window
has 500 MHz bandwidth and 15 channels while each narrow-band window has 31 MHz
bandwidth and 63 channels. The narrow-band high spectral resolution is 0.49 MHz
per channel, which provides sufficient sensitivity to resolve the observed
transitions and lower the line confusion effect to detect the compact emission
from large molecules. The amplitude and phase calibrations were accomplished by
observing the gain and flux calibrators. The overall flux calibration
uncertainty is 10-15\%. The CARMA flux uncertainty discussed in the text is the
statistical uncertainty and does not account for any amplitude calibration
error. In this paper, we will use the style of showing 1$\sigma$ statistical
uncertainties in parenthesizes after the derived values. During data reduction,
pointing and system temperatures were checked to monitor data quality.

We observed the $10_{*,10}-9_{*,9}$ $E$ and $A$ CH$_3$COOH lines at
111.507270(20) and 111.548533(20) GHz \citep{ilyu08} in two narrow band windows.
Based on previous observations \citep{remi02,remi03}, they are unblended and
have similar line strengths. In addition to CH$_3$COOH, we included two
HCOOCH$_{3}$ and three CH$_3$CH$_2$CN lines in the other narrow-band  windows.
Table \ref{tbl-mol} summarizes the observed molecules and their transitions.  
All spectral line data were taken from the Spectral Line Atlas of Interstellar
Molecules (SLAIM) (F. J. Lovas, private communication)\footnote{available at 
http://www.splatalogue.net \citep{remi07}}, the Cologne Database for Molecular
Spectroscopy \citep{mull05} and the appropriate laboratory references listed
therein.

\section{RESULTS}

The observational data were reduced with the MIRIAD package \citep{sault95}. The
final data presented in the figures were Hanning-smoothed over 3 channels and
continuum-subtracted.  The observations in B and C configurations were combined
using natural weighting to obtain the best sensitivity and synthesized beams of
$\sim$2$\arcsec$.  We present the observations of each source as both maps  of
the molecular emission and the spectra at the peak of the emission (Figures 1
through 7). The continuum emission maps shown were made from channels contained
in the  wideband windows which were free from line emission. All of the spectra
from each source are shown in rows and columns. In each spectral figure, column
(a) displays window 2 (rest frequency=107.54375 GHz); column (b), window 3 (rest
frequency=107.59404 GHz); column (c), window 6 (rest frequency=111.50727 GHz);
and column (d), window 5 (rest frequency=111.54853 GHz). In row 1, the observed 
spectra (Hanning smoothed over 3 channels) are overlaid on the modeled spectral
line data represented by Gaussians (red trace). In row 2, the modeled spectral
line data from row 1 are convolved with the spectral resolution of the original
observational data and co-added with 24 mJy/beam random noise. Finally, row 3
presents the residuals from the modeled data subtracted from the observational
data (i.e row 1 minus row 2). The maps and spectra are discussed below.

\subsection{G19.61-0.23}

\begin{figure}
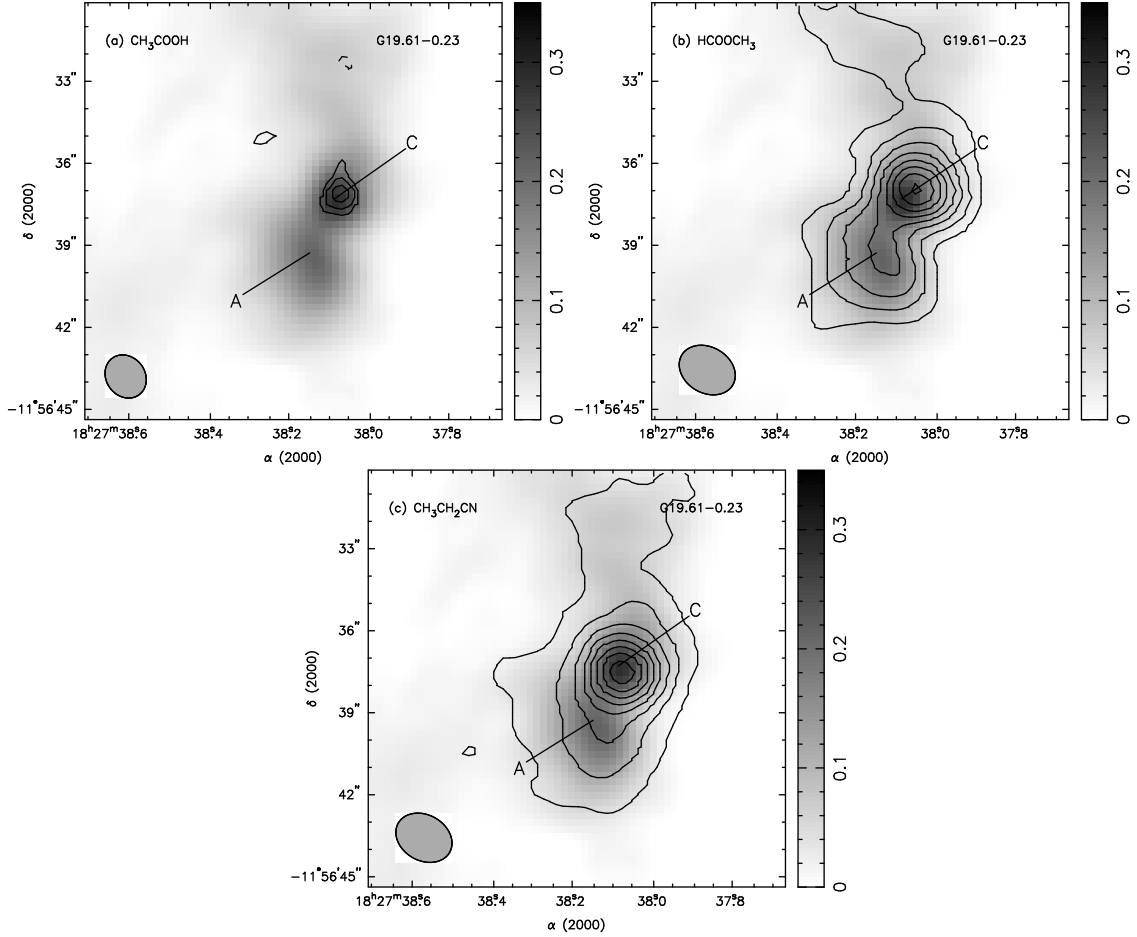

\begin{center}
\includegraphics[angle=270,scale=0.35]{g19aca_map.ps}
\includegraphics[angle=270,scale=0.35]{g19mef_map.ps}
\includegraphics[angle=270,scale=0.35]{g19etcn_map.ps}
\caption{G19 maps.  (a) CH$_3$COOH contours overlaid on the gray continuum map.
The  continuum emission maps shown were made from channels contained in the 
wideband windows which were free from line emission. The contour levels are -3,
3, 4 and 5 $\times$ $\sigma=24$ mJy beam$^{-1}$. The gray scale unit is Jy
beam$^{-1}$. The synthesized beam size is 1\farcs65$\times$1\farcs42
(P.A.=46.3\degr), shown in the bottom left corner. The continuum beam size is
1\farcs58$\times$1\farcs45 (P.A.=36.4\degr). In the continuum map, there are two
main components, A and C. The CH$_3$COOH emission is mainly overlaid on
component C. (b) HCOOCH$_3$ contours overlaid on the gray continuum map. The
contour levels averaged over 3 channels are 3, 6, 9, 12, 15, 18, 21, 24, 27 and
30 $\times$ $\sigma=20$ mJy beam$^{-1}$. The synthesized beam size is
2\farcs16$\times$1\farcs67 (P.A.=59.6\degr). (c) CH$_3$CH$_2$CN contours
overlaid on the gray continuum map. The contour levels averaged over 6 channels,
are 3, 9, 15, 21, 27, 33, 39, 45, 51, 57 and 63 $\times$ $\sigma=18$ mJy
beam$^{-1}$. The synthesized beam size is 2\farcs16$\times$1\farcs67
(P.A.=60.3\degr).}
\label{g19map}
\end{center}
\end{figure}

\begin{figure}
\begin{center}
\includegraphics[angle=270,scale=0.7]{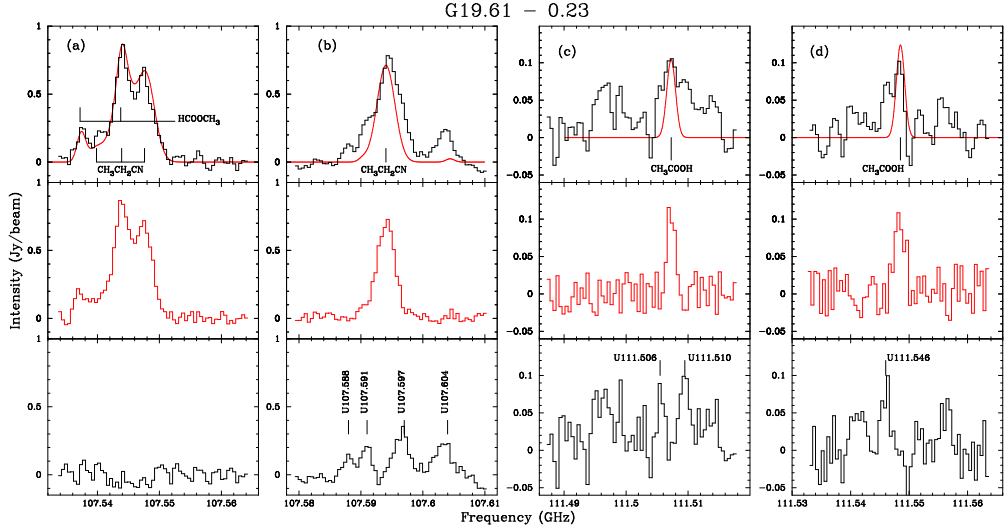}
\caption{The observed spectra of G19.61-0.23. The first column (a) presents two
HCOOCH$_3$ lines, $9_{2,8}-8_{2,7}$ $E$ and $9_{2,8}-8_{2,7}$ $A$; and four
CH$_3$CH$_2$CN lines, $12_{4,9}-11_{4,8}$, $12_{4,8}-11_{4,7}$,
$12_{11,2}-11_{11,1}$ and $12_{11,1}-11_{11,0}$. The $12_{11,2}-11_{11,1}$ and
$12_{11,1}-11_{11,0}$ lines are not spectrally resolved. The v$_{LSR}$ range is
-15 to 69.4 km s$^{-1}$. The second column (b) contains only one known line,
CH$_3$CH$_2$CN $12_{3,10}-11_{3,9}$. The v$_{LSR}$ range is -3.4 to 81 km
s$^{-1}$. The third column (c) contains the CH$_3$COOH $10_{*,10}-9_{*,9}$ $E$
line and the fourth column (d) contains the CH$_3$COOH $10_{*,10}-9_{*,9}$ $A$
line. The rest frequencies of the windows are 107.54375, 107.59404, 111.50727
and 111.54853 GHz, respectively. The v$_{LSR}$ ranges are 11.3 to 92.7 and -1.8
to 79.6 km s$^{-1}$ for column (c) and (d), respectively. The first row presents
the observed spectra (Hanning smoothed over 3 channels) overlaid on modeled
spectral line data represented by Gaussians. The second row shows the modeled
spectral line data from the first row convolved with the spectral resolution of
the original observation and with random noise added to the spectrum. The value
of the random noise is obtained with the line-free continuum spectrum in the
windows. Finally, the third row  shows the residuals from the modeled data in
the second row subtracted from the observational data in the first row.}
\label{g19spec}
\end{center}
\end{figure}

One of the most relevant discoveries in this paper is the new detection of
CH$_3$COOH toward G19 that highlights the importance of interferometric
observations in the detection of weak spectral signatures confined in compact 
emission regions. To illustrate the distribution of the CH$_3$COOH emission
toward G19, we have summed the intensity of the CH$_3$COOH lines to produce a
combined CH$_3$COOH contour map which is overlaid on the continuum map in
grayscale in Figure \ref{g19map}(a). Both the $10_{*,10}-9_{*,9}$ $E$ and 
$10_{*,10}-9_{*,9}$ $A$ lines were detected in the two CH$_3$COOH spectral
windows (Figures \ref{g19spec}(c) and (d)) at the canonical
$v_{LSR}$ of 40 km s$^{-1}$ for G19. While some weak unidentified lines may
appear in the windows after Hanning smoothing, the relative line strengths of
the two CH$_3$COOH transitions are roughly equal as predicted by rotational
spectroscopy. G19 consists of several high density components of continuum
emission between 2 cm and 3 mm wavelengths \citep{furu05}. In this case, the
CH$_3$COOH emission is located near the center of component C and the CH$_3$COOH
emission region has an effective scale size of 2\arcsec.

In addition to CH$_3$COOH, we have observed transitions of HCOOCH$_{3}$ and
CH$_3$CH$_2$CN. Two transitions of HCOOCH$_{3}$, $9_{2,8}-8_{2,7}$ $A/E$ and
three transitions of CH$_3$CH$_2$CN, $12_{4,9}-11_{4,8}$, $12_{4,8}-11_{4,7}$,
and $12_{3,10}-11_{3,9}$ are shown in Figures \ref{g19spec}(a) and (b). All
transitions are at a $v_{LSR}$=40 km s$^{-1}$. The statistical significance of
three possible unidentified lines will be discussed in Section 5.5. The
HCOOCH$_{3}$ map (Figure \ref{g19map}(b)) and CH$_3$CH$_2$CN map (Figure
\ref{g19map}(c)) show emission that peaks primarily near component C, like the
CH$_3$COOH emission, yet with a southward projection toward component A that was
not seen in the lower resolution BIMA observations \citep{remi04}. Table
\ref{tbl-g19} summarizes the fitting results of the detected transitions towards
G19. In our fitting routine, we assume that the rest frequencies of each
molecular transition is fixed to the $v_{LSR}$ of the source and the
HCOOCH$_{3}$  transition that is blended with the CH$_3$CH$_2$CN transition
shares the same linewidth and peak intensity as the unblended transition. Also,
shown are the line widths and intensities of the model data for a direct
comparison.

\subsection{G29.96-0.02}

\begin{figure}
\begin{center}
\includegraphics[angle=270,scale=0.4]{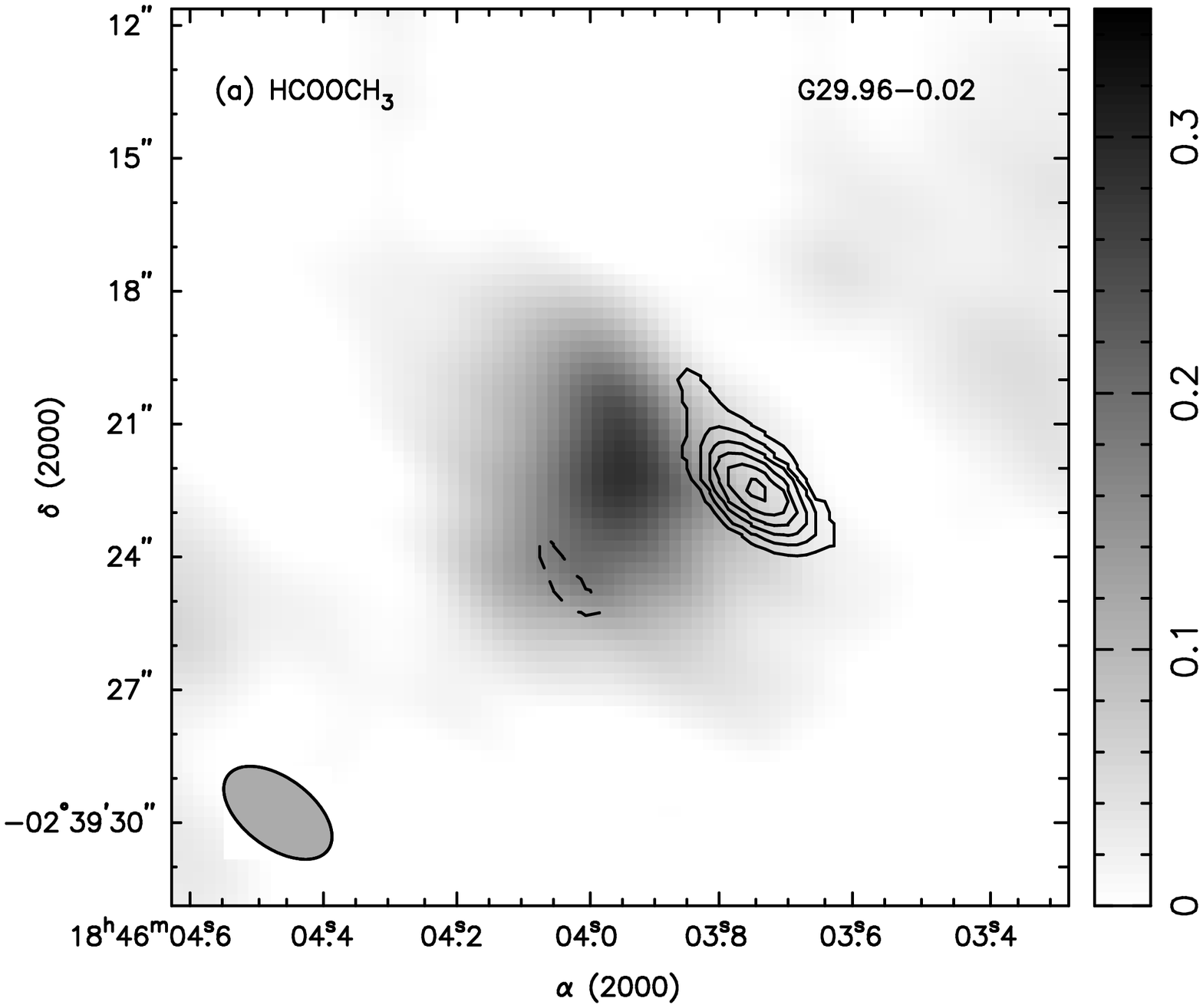}
\includegraphics[angle=270,scale=0.4]{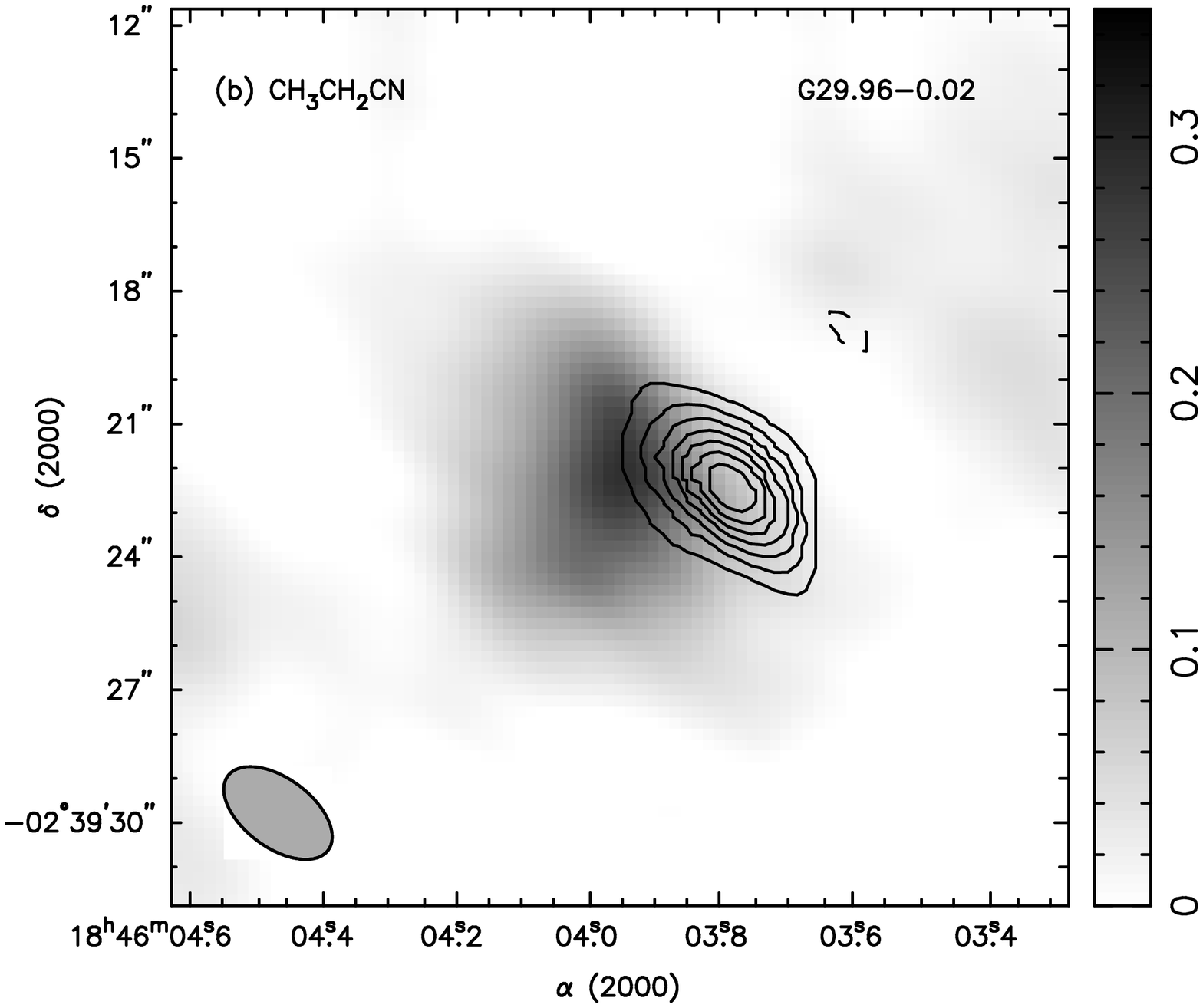}
\caption{G29 maps. (a) HCOOCH$_3$ contours overlaid on the gray continuum map.
The contour levels averaged over 3 channels are -3, 3, 4, 5, 6, 7 and 8 $\times$
$\sigma=21$ mJy beam$^{-1}$. The synthesized beam size is
2\farcs81$\times$1\farcs56 (P.A.=53.2\degr), shown in the bottom left corner.
(b) CH$_3$CH$_2$CN contours overlaid on the gray continuum map. The contour
levels averaged over 6 channels, are -3, 3, 5, 7, 9, 11, 13 and 15 $\times$
$\sigma=16$ mJy beam$^{-1}$. The synthesized beam size is
2\farcs82$\times$1\farcs56 (P.A.=53.5\degr).}
\label{g29map}
\end{center}
\end{figure}

\begin{figure}
\begin{center}
\includegraphics[angle=270,scale=0.7]{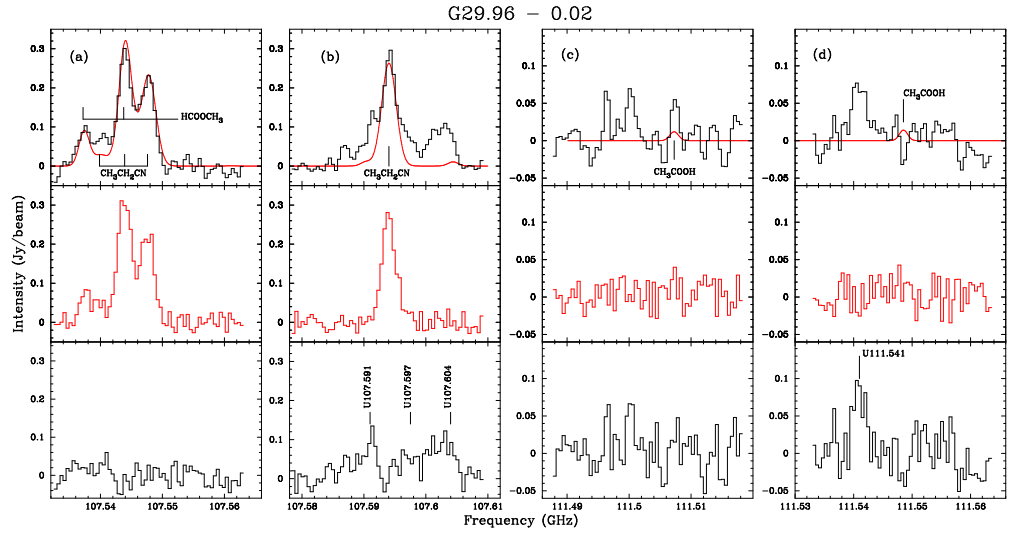}
\caption{The observed spectra of G29.96-0.02. The arrangement of frequency and
windows is the same as Figure \ref{g19spec}. The v$_{LSR}$ ranges are
44.6 to 129, 56.2 to 140.6, 68.8 to 150.2 and 55.7 to 137.1 km
s$^{-1}$, respectively.}
\label{g29spec}
\end{center}
\end{figure}

We have detected HCOOCH$_3$ and CH$_3$CH$_2$CN toward G29. Figures
\ref{g29map}(a) and (b) show the distribution of HCOOCH$_{3}$  and
CH$_3$CH$_2$CN emission toward the G29 region, respectively.  In Figure
\ref{g29spec}, we clearly detected the two blended pairs of transitions of
HCOOCH$_3$ $9_{2,8}-8_{2,7}$ $E$/$A$ and CH$_3$CH$_2$CN $12_{4,9}-11_{4,8}$ and
$12_{4,8}-11_{4,7}$. The spectrum is very similar to that of G19 (Figure
\ref{g19spec}(a)). The CH$_3$CH$_2$CN $12_{3,10}-11_{3,9}$ transition and
possible U lines in Figure \ref{g29spec}(b), also have similar profiles to those
in Figure \ref{g19spec}(b). 

G29 consists of two main continuum peaks at 1.2 cm, which trace the UC\ion{H}{2}
region, and at 862 $\mu$m, which trace the cloud core \citep{beut07}. The cloud
core is the location of the molecular emission; however, the emission is not 
bright enough to contribute significantly at 3 mm wavelengths. In Figure
\ref{g29map} the continuum emission is dominated by the UC \ion{H}{2} region.

Figure \ref{g29spec}(c) and (d) shows the two CH$_3$COOH spectral windows toward
G29. Unlike the data from G19, there are no statistically significant features
at the CH$_3$COOH rest frequencies. There is a 1$\sigma$ feature that may be due
to the $10_{*,10}-9_{*,9}$ $E$ transition of CH$_3$COOH, but the corresponding
$10_{*,10}-9_{*,9}$ $A$ transition is noticeably absent. In the discussion, we
will investigate these passbands more closely with respect to the modeled data.
Table \ref{tbl-g29} summarizes the fitting results of the detected transitions
towards G29. The fitting criteria and model predictions are presented in the
same manner as in Table \ref{tbl-g19}.

\subsection{IRAS 16293-2422}

\begin{figure}
\begin{center}
\includegraphics[angle=270,scale=0.4]{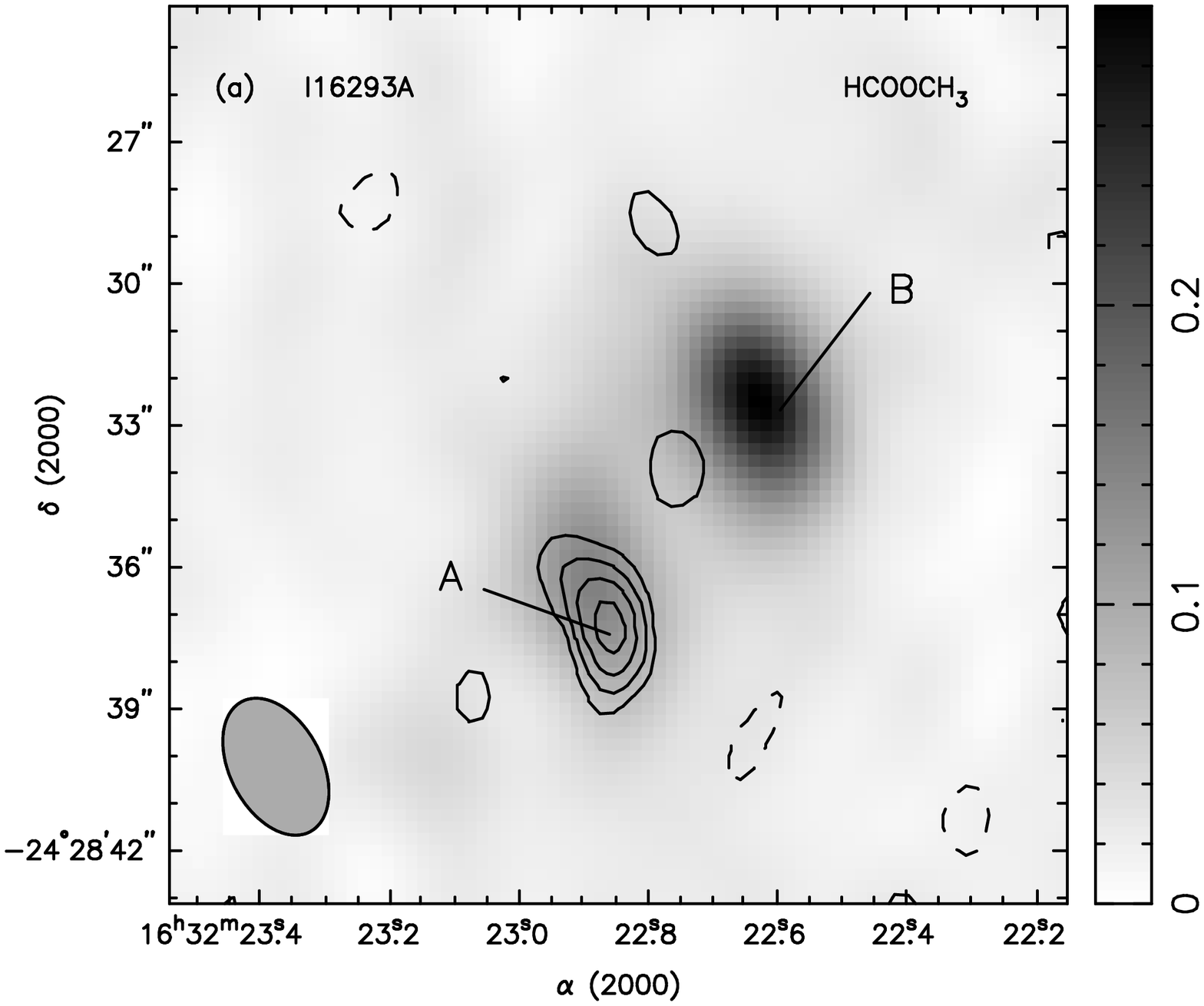}
\includegraphics[angle=270,scale=0.4]{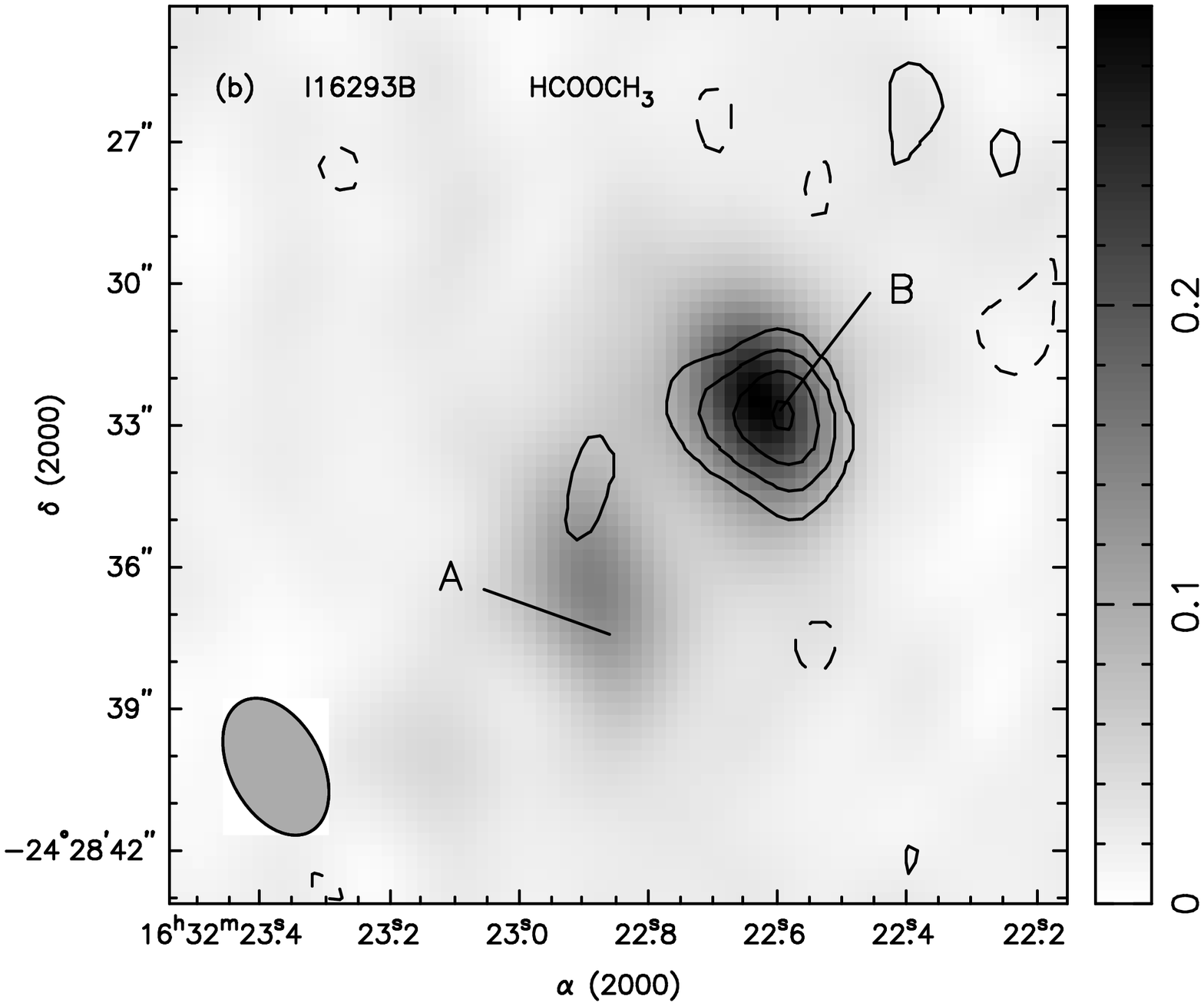}

\caption{The HCOOCH$_3$ contour maps are made individually for I16293A and B
(denoted as A and B on the maps) due to the different linewidths. The  contours
are overlaid on the gray continuum map of IRAS 16293-2422. (a) The contours
located at I16293A are made with the $9_{2,8}-8_{2,7}$ $E$ line averaged over
six channels. The contour levels are -2, 2, 3, 4 and 5 $\times ~\sigma=21$ mJy
beam$^{-1}$. (b) The contours located at I16293B are made with one channel of
the $9_{2,8}-8_{2,7}$ $E$ line. The contour levels are -2, 2, 3, 4 and 5 $\times
~\sigma=25$ mJy beam$^{-1}$.}

\label{i16293map}
\end{center}
\end{figure}

\begin{figure}
\begin{center}
\includegraphics[angle=270,scale=0.7]{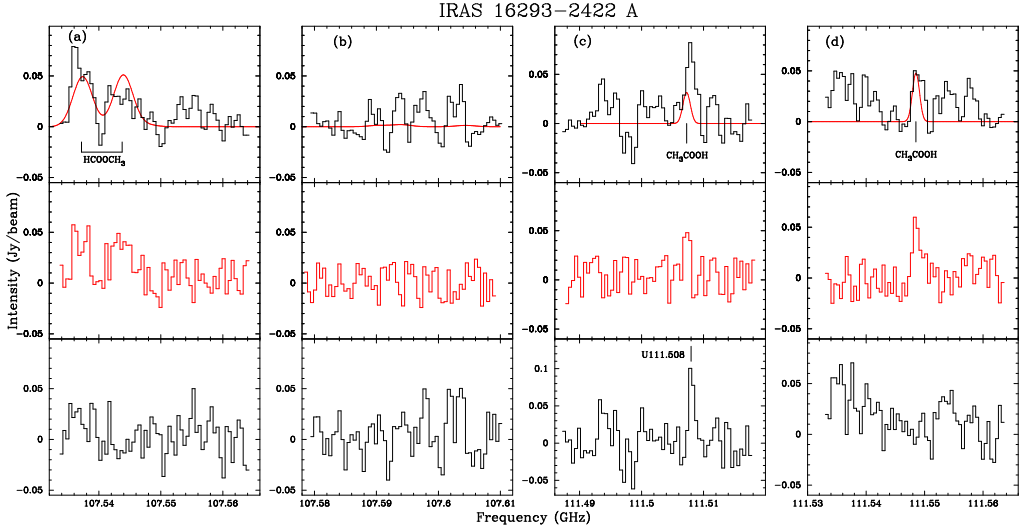}
\caption{The observed spectra of I16293A. The arrangement of frequency and
windows is the same as Figure \ref{g19spec}. The column v$_{LSR}$ ranges are
44.6 to 129, 56.2 to 140.6, 68.8 to 150.2 and 55.7 to 137.1 km s$^{-1}$,
respectively.}
\label{spec_i16293a}
\end{center}
\end{figure}

\begin{figure}
\begin{center}
\includegraphics[angle=270,scale=0.7]{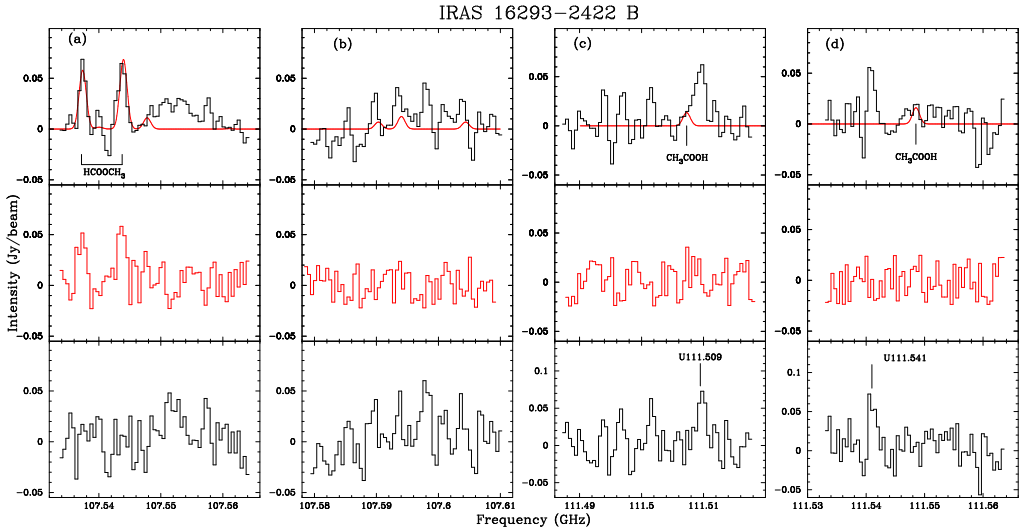}
\caption{The observed spectra of I16293B. The arrangement of frequency and
windows is the same as Figure \ref{spec_i16293a}. The column v$_{LSR}$ ranges
are -54.4 to 30, -42.7 to 41.6, -24.8 to 56.6 and -36.8 to 44.6 km s$^{-1}$,
respectively.}
\label{spec_i16293b}
\end{center}
\end{figure}

Unlike the spectra from \citet{cazaux03}, we resolved IRAS 16293-2422 into
I16293A and B with our 3\farcs1$\times$ 1\farcs9 beam (see Figure
\ref{i16293map}). Therefore, beam dilution in this observation is minimized. 
Figure \ref{spec_i16293a} and \ref{spec_i16293b} shows the two CH$_3$COOH
spectral windows toward I16293A and I16293B, respectively. Toward I16293A, we
detect weak emission features that may be due to transitions of CH$_3$COOH but
they are right at the 2$\sigma$ detection limit. Toward I16293B, as in G29,
there are no statistically significant features at the CH$_3$COOH rest
frequencies.  There are 1$\sigma$ features that may be due to the
$10_{*,10}-9_{*,9}$ $A/E$ transitions in these Hanning smoothed data, but
nothing convincing beyond that level. In $\S$5.5 we evaluate whether the 
features toward I16293A may be due to CH$_3$COOH.

We have also observed two new transitions of HCOOCH$_{3}$. The line profiles of
HCOOCH$_{3}$ shown in Figure \ref{spec_i16293a} toward I16293A are wider and may
be suffering from self absorption \citep{remi06} more than the line profiles 
shown in Figure \ref{spec_i16293b} toward I16293B, which are much more narrow.
This suggests that I16293A and I16293B may have distinguishable molecular
dynamics. Figure \ref{i16293map} (a) and (b) show the distribution of
HCOOCH$_{3}$ emission with respect to the 3 mm continuum emission toward the
I16293A \& B regions, respectively. Table \ref{tbl-i16293} summarizes the
fitting results of the  detected transitions towards IRAS 16293-2422. The
fitting criteria and model predictions are presented in the same manner as in
Table \ref{tbl-g19}.

\section{ANALYSIS \& DISCUSSION}

\subsection{Molecular Column Densities} 

Molecular column density is one essential physical parameter needed to constrain
chemical models (see e.g. Garrod et al. 2008 and references therein). In order
to  determine the column densities of each molecular  species of interest, we
assume that each region has uniform physical conditions, that the populations of
the energy levels can be characterized by a Boltzmann distribution, and finally,
that the emission is optically thin.  Assuming that the  molecular species is in
local thermodynamic equilibrium (LTE) and low  optical depth, the total beam
averaged column density is
\begin{equation}
N_T (\mathrm{cm^{-2}})=2.04\times10^{20} \it \frac{\int I(\mathrm{Jy~beam^{-1}})
d\upsilon(\mathrm{km~s^{-1}})Q_re^{E_u(K)/T_r}}{\Omega_b(\mathrm{arcsec^2})
\nu^{3}(\mathrm{GHz^3})<S\mu^{2}>(\mathrm{debye^2})},
\label{eqn_Ntot}
\end{equation}
where $\Omega_b$ is the solid angle of the beam , $\int I_{\upsilon}d\upsilon$
is the integral of the line intensity over velocity, $\nu$ is the spectral line
frequency, $<S\mu^{2}>$ is the line strength parameter, $Q_r$ is the rotational
partition function, $T_r$ is the rotational temperature, and $E_u$ is the upper
level energy of the transition \citep{miao95}. In each case, we assume the
source emission fills the synthesized beam of the observations.  All of the line
parameters used for the analysis are given in Table 2.

A least-square fitting routine with Gaussian functions was used to determine 
the linewidths ($\Delta v$), peak intensities ($\Delta$I) and $\int I d\upsilon$
of each observed spectral line toward each source.  All least-square fitting was
done using the  standard packages contained in the $Mathematica$ software
package.  Using the spectroscopic  parameters from Table 2 and by varying the
temperature and column density over a wide  range of values (e.g. the
temperature range investigated was from 10-500 K and the column  density range
was from 10$^{12}$ - 10$^{18}$ cm$^{-2}$ for each molecular species), Equation
(1) gave a predicted integrated line intensity for each spectral line observed
in our data. These predictions of integrated line intensity were then compared
to the measured values and, using a least-squares fitting routine, a best fit
temperature and column density was found for each molecule toward each source.
However, there were some sources where an independent fit of the temperature did
not converge.  These sources are described in more detail in the  following
sections. Columns 3 and 4 of Table 3, 4 and 5 give the fitted line parameters of
the  observed peak intensity and line width. Columns 5 and 6 give the modeled
peak intensity  based on the least square fitting routine and the fitted line
width from the observed data,  respectively.  Once all the relevant line
parameters have been determined, a model spectrum  with the appropriate amount
of random noise added and convolved with the spectral resolution of the
observation can then be produced and be directly compared with the observed
dataset (see $\S$5.5).

The rotational temperature $T_r$ generally is representative of the physical 
kinetic temperature in regions where the spatial density is larger than the
critical density of the molecule in question. The assumption in this work is
that  the densities of the HMCs investigated are expected to be high enough for
this to be true for these molecules.  The rotational temperature is often found
by observing several different transitions of a molecular species over a range
of $E_u$ values.  However, given the limited bandwidth and low number of
transitions observed in our data, it was often difficult to accurately constrain
the temperature and column density of our three observed molecules toward these
regions \citep[e.g.,][]{snyder05}. Therefore, when the least squares fit failed
to give a reasonable temperature for a given molecular species, we adopted a
rotational temperature from previous observations by other investigators toward
these sources at similar spatial resolution and were then able to fit for the
total beam averaged column densities or associated upper limits if no
transitions were detected beyond the statistically significant 1$\sigma$
limit.  It was also the case that if the fitting procedure was given too many
free parameters, a constraint was needed on one or more or else it would not
converge (see the G29 discussion below).  To properly constrain the temperature
of these regions over  a variety of molecular species, observations of
additional transitions are  necessary and complete spectral line surveys may be
required \citep[see e.g.][and references therein]{font07,frie04}.

For the G19 region, the least square fitting routine determined a rotational
temperature of CH$_3$CH$_2$CN of T$_{rot}$=161(58) K, and a beam averaged column
density of  $N_T$=6(3)$\times$10$^{16}$ cm$^{-2}$.  The errors on all fits are
1$\sigma$. The routine could not find a reasonable fit for the temperature and
column density  for either HCOOCH$_3$ or CH$_3$COOH presumably because the
detected transitions  were too close in energy.  Therefore, because the emission
regions of each molecular species  around G19 are nearly co-spatial (Figure
\ref{g19map}), we adopted the same rotational temperature for HCOOCH$_3$ and
CH$_3$COOH as found from the CH$_3$CH$_2$CN fit.  Assuming a  T$_{rot}$=161 K
for HCOOCH$_3$, we find a best fit to the column density of
$N_T$=9(2)$\times$10$^{16}$ cm$^{-2}$ and for CH$_3$COOH, we find a best fit to
the  column density of $N_T$=2(1)$\times$10$^{16}$ cm$^{-2}$.  Given the minimum
and maximum values of the measured column density, we determine an abundance
ratio range of N$_{HCOOCH_3}$/N$_{CH_3COOH}$=3-11. Table \ref{tbl-g19} gives the
fitted intensities of each of the detected spectral features compared to the
predicted intensities determined from Equation (1) for G19.

For the G29 region, the least square fitting routine determined a rotational
temperature of CH$_3$CH$_2$CN of T$_{rot}$=107(49) K, and a beam averaged column
density of  $N_T$=1.1(7)$\times$10$^{16}$ cm$^{-2}$ only if we adopted a
rotational temperature of 150 K, determined from \citet{olmi03} and supported by
the observations of \cite{font07} and \cite{beut07}, for the transitions of
HCOOCH$_3$.  This temperature was also set to determine the upper limit for the
column density of CH$_3$COOH. Assuming a  T$_{rot}$=150 K for HCOOCH$_3$, we
find a best fit to the column density of $N_T$=4(1)$\times$10$^{16}$ cm$^{-2}$
and for CH$_3$COOH, we find an upper limit to the column density of
$N_T$$<$9$\times$10$^{14}$ cm$^{-2}$ for a relative abundance ratio upper limit
of N$_{HCOOCH_3}$/N$_{CH_3COOH}$$>$50.  Table \ref{tbl-g29} gives the fitted
intensities of each of the detected spectral features compared to the predicted
intensities determined from equation (1) for G29.  We also note the detection of
another feature shown in Figure \ref{g19spec} near 107.540 GHz that is most
likely due to two high energy ($>$110 K) spectral lines of CH$_3$CH$_2$CN.  The
discussion of weak spectral features  near the 1$\sigma$ noise level will be
given in section $\S$5.5.

A recent observation of the G19 and G29 regions with the IRAM 30-m telescope
\citep{font07} covered our two acetic acid line frequencies (Figure
\ref{spec_30m}). However, with a 28\arcsec~beam, the lines detected are under
the noise level of 26 mK. These lines are brighter than those observed  by
\cite{cazaux03}, but the beam dilution is severe. The high resolution capability
of CARMA is indeed critical for searching for acetic acid and large molecule
research.

Finally, toward IRAS 16293-2422, the least square fitting routine could not
determine a rotational temperature for the detected transitions of HCOOCH$_3$
presumably for the same reason as described toward G19. Thus, we adopted a
rotational temperature of 62 K determined from \citet{bott04}. In this case,
we find a best fit to  the column density of  HCOOCH$_{3}$ of
$N_T$=1.5(3)$\times$10$^{16}$ cm$^{-2}$  toward region A and
6(1)$\times$10$^{15}$ cm$^{-2}$ toward region B. Since there was no clear
detection of CH$_3$CH$_2$CN  beyond the 1$\sigma$ detection limit toward either 
region A or B, based on the noise level in each of the  passbands containing 
those lines, we determined an upper limit to the total beam averaged column
density  of CH$_3$CH$_2$CN to be $<$6$\times$10$^{13}$ cm$^{-2}$ for region
A; and $<$3$\times$10$^{14}$ cm$^{-2}$ toward region B.  For CH$_3$COOH, the
2$\sigma$ detection toward region A gives a total beam averaged column density 
of $\sim$1.6$\times$10$^{15}$ cm$^{-2}$ and for region B, we determine and
upper  limit of $<$6$\times$10$^{14}$ cm$^{-2}$.  Table \ref{tbl-i16293}
summarizes the  fitting results of the HCOOCH$_{3}$ transitions compared to the
model intensities  determined from Equation (1).

\subsection{A Proxy for the Detection of CH$_{3}$COOH Sources}

We have successfully detected CH$_{3}$COOH toward G19. Following Sgr B2(N-LMH), 
W51e2 and G34.3+0.15, G19.61-0.23 is the fourth high mass hot core that is a
source of CH$_3$COOH. We compare the CH$_3$COOH column densities and the
abundance ratios to HCOOCH$_3$ of all detected CH$_3$COOH hot cores in Table
\ref{tbl-aca_cmp}. The values are generally consistent. Therefore, the column
densities and ratios can be used to constrain hot core chemical models assuming
the co-spatial dependence of HCOOCH$_3$ and CH$_{3}$COOH remains.   This has yet
to be tested at extremely high spatial resolution. In addition, a
consistency can be found between all the sources containing CH$_3$COOH emission by comparing
the relative line strengths between the detected transitions of HCOOCH$_3$ and
CH$_3$CH$_2$CN to the CH$_3$COOH emission.  Taking a ratio between measured (or
predicted intensities) of the 107.594 GHz line of CH$_3$CH$_2$CN to the
111.507 GHz line of CH$_3$COOH, we find that this ratio is between 5.5 and 6 for
those sources where CH$_3$COOH is detected. Therefore, given the line intensity
of 0.21 Jy/beam for the 107.594 GHz line of CH$_3$CH$_2$CN toward G29.96-0.02,
the expected line intensity for the 111.507 GHz line of CH$_3$COOH is $\leq$0.04
Jy/beam.  This intensity is right at the 1$\sigma$ detection limit of the
current CARMA observations.  Therefore, in addition  to the criteria for the
detection of CH$_3$COOH outlined in \citet{remi03}, we now have a good
proxy for the detectability of the 111 GHz CH$_3$COOH transitions relative to
the detected line strengths of the HCOOCH$_3$ and CH$_3$CH$_2$CN transitions
near 107 GHz. Further observations of new sources of  HCOOCH$_3$ and
CH$_3$CH$_2$CN that fit the criteria set by \citet{remi03} and higher
S/N observations of sources like G29, are necessary to validate this
hypothesis.  

\subsection{CH$_{3}$COOH in IRAS 16293-2422?}

High spatial resolution maps and spectra of IRAS 16293-2422 clearly reveal that
the morphologies of the emission of complex molecules are clearly not co-spatial
and that there is no simple separation of O and N-bearing ``chemical
differentiation''  \citep{guelin08}.  \citet{blak87} and \citet{rodg01} have
suggested the difference in the morphologies of the emission of complex
molecules in these regions reflects the timescales of chemical evolution.
Predominantly N-bearing species (e.g.\ CH$_3$CH$_2$CN) represent the later
stage of chemical evolution while  predominantly O-bearing species (e.g.\
HCOOCH$_3$) represent the earlier stage.  Their models can estimate the ages of
hot cores by comparing this difference between the O and N  morphology of
emission.  However, to date, no direct observation can accurately  measure hot
core ages, so the uncertainty of these estimates is large.

IRAS 16293-2422 also shows clear distinctions in the morphology of molecular
emission \citep{chan05,remi06}.  In fact, the relative distributions of
molecular species will undoubtedly give insight into their formation chemistry. 
It may also turn out that the relative abundances of molecular species and the
overall population of  energy levels of the detected transitions of a given
molecule may be determined  more by the formation chemistry and not on the local
thermodynamic environment.  To test this hypothesis will require high spectral
and spatial resolution observations  of a group of species that may form via a
common formation pathway.

IRAS 16293-2422 has been reported as the first low-mass CH$_3$COOH source by
\citet{cazaux03} using the IRAM 30-m telescope, but the one detected line was
likely blended. While IRAS 16293-2422 is not resolved in their work (beamsize of
28\arcsec),  we have resolved it into I16293A and B with a separation of less
than 5\arcsec. Even though their observation used different  transitions, we
should detect our CH$_3$COOH lines in our synthesized beam of
3\farcs1$\times$1\farcs9 using their flux estimates.  Although at a lower flux
than expected from their results, we do detect two CH$_3$COOH transitions toward
region A but only at the 2$\sigma$ limit.  We estimate a column density of
$\sim$ 1.6 $\times$ 10$^{15}$ cm$^{-2}$, and \citet{cazaux03} estimated an upper
limit of 2.5 $\times$ 10$^{15}$ cm$^{-2}$, which is similar.

Furthermore, there is no evidence beyond the 1$\sigma$ levels of any emission of
CH$_3$COOH toward region B. In addition, while we have observed  HCOOCH$_3$ in
I16293A and B, we did not detect the corresponding CH$_3$CH$_2$CN emission
features. The detection of CH$_3$COOH toward G19 (and possibly toward I16293A)
and the other hot cores have demonstrated that a diversity of abundant large O
and N-bearing species are strongly correlated with CH$_3$COOH. One possibility
is that the mix of O and N-bearing species may serve as an efficient chemical
network for CH$_3$COOH and other large molecules. Thus, they need to wait for
large N-bearing molecules to form first before they can be efficiently produced.
Another possibility is that the formation of large O and N-bearing species is
simply time dependent, and the gas-phase CH$_3$COOH concentration  just by
coincidence starts to increase along with large N-bearing species.  Given the
criteria set above, it is intriguing that CH$_3$COOH is detected toward I16293A
given there was no clear evidence of any emission from CH$_3$CH$_2$CN.  Further
higher sensitivity observations are once again necessary to confirm this
detection and determine the evolutionary stage of the I16293A and B cores.

\subsection{CH$_{3}$COOH Formation}

Both gas phase and grain surface chemistry undoubtedly play important roles in
hot cores \citep[e.g.][]{hase92,garr06,garr08}. Large molecule abundances can be
enhanced enormously by grain surface chemistry while current models of gas phase
chemistry alone are not sufficient to match observations.  \citet{garr06} have
demonstrated that radicals in the icy mantles on grain surfaces can move around
as the temperature gradually increases from 10 to 200 K in star forming regions.
Grain surfaces become efficient environments for large molecules to form.
However, the details of CH$_3$COOH formation in hot cores still remain unclear.
The correlation between CH$_3$COOH and large N-bearing molecules may suggest
that some N-bearing molecules act as catalysts to the CH$_3$COOH formation. The
chemical and physical evolution timescales of star formation lacks direct
observational evidence to confirm chemical models. Without the support of
observation, it is difficult to establish solid chemical models. \citet{garr08}
have reported their new chemical model for CH$_3$COOH in hot cores. Grain
surface reactions have been enhanced to match observations. They show the time
scales of hot cores warming up affect the abundances of the secondary radical
CH$_3$CO, which can react with OH to form CH$_{3}$COOH. This effect can also
cause the dissimilar abundances of structural isomers, which is consistent with
our survey.

\subsection{Identifying Weak Spectral Features Using Model Spectra.}

When CH$_3$COOH was first detected \citep{meh97}, it emphasized the importance
of interferometry for the study of astrochemistry in hot cores. Star forming
regions generally contain many regions of compact molecular emission.  Although
G19 has a complex structure, including extended and compact components
\citep[see][]{furu05}, the CH$_3$COOH detection toward G19 was conducted with a
high resolution beam, 1\farcs9$\times$1\farcs6, which is the smallest used in a
CH$_{3}$COOH survey to date. The three observed molecules in G19, CH$_3$COOH,
HCOOCH$_3$ and CH$_3$CH$_2$CN, are mainly located at the same component, which
agrees with the CH$_3$CH$_2$CN observation by \citet{furu05}. As we continue
extending CH$_{3}$COOH surveys to smaller hot cores, the required resolution of
observations will be higher. 

Higher resolution observations also have the advantage of reducing the effect of
line confusion present in a spectrum.  Also, if the source size of the emitting
region is well coupled to the synthesized beam of the array, weak spectral
features that are lost due to beam dilution effects from single dish
observations can be significantly resolved out of the noise. Thus, many new
molecule detections over the last several years have used both a combination of
single dish and array observations to confirm the detection
\citep[e.g.,][]{bell08,bell09}. In addition, these detections routinely use
model spectra based on the laboratory and calculated rotational spectroscopy of
energy levels and line strengths for molecular transitions, the physical
conditions of the emitting regions including temperature and density and
finally, the characteristics of the telescope including forward beam efficiency
and field of view  (i.e. beam size).  Given the model spectra, the weak
transitions of molecular species appear statistically significant in the
observed spectrum. However, the model data have near infinite spectral
resolution compared to the observed data and do not include an estimate of the
noise level in the passband.  These types of models are shown in the spectral
line data presented in this work (red Gaussian traces in the first row of spectra
in Figures 2, 4, 6 and 7).

In order to correctly ascertain whether a weak feature in the observed spectrum
is due to  a calculated line of a molecular species, we advocate taking the
model data and convolving it with the spectral resolution of the instrument and
adding the appropriate amount of random noise to the resultant spectrum to give
a more accurate representation of  the data collected by the telescope.   The
noise level is based on the measured noise of line free channels in the observed
data.  These modeled data are then subtracted from the observed data to form a
residual  spectrum.  From these residuals, with the caveat of an accurate model
of the physical conditions of the source, one can identify the remaining weak
spectral features and determine if features close to the 1-2$\sigma$ level are
in fact statistically significant spectral line features.

Based on the residual passbands shown in the last row of Figures 2, 4, 6 and 7,
we include in Tables 3, 4 and 5 the possibility of unidentified lines (i.e.
U-lines) in the data toward G19, G29 and IRAS 16293-2422. In addition, given the
spectral line fit (red Gaussian traces) toward IRAS 16293-2422, there is a
suggestion that the weak emission features shown toward I16293B may in fact be
due to CH$_3$COOH.  However, when this model spectrum is convolved with the
noise level in the passband, these weak features completely disappear and there
are definitely no statistically significant feature seen in the residual
passbands. As a result, while we claim that the  emission features that are
coincident with the CH$_3$COOH transitions toward I16293A are in fact from
CH$_3$COOH, these features are only  barely above the noise level so further
high sensitivity observations are  needed for a definitive detection.

It is our recommendation that if model spectra are going to be the new proxy for
identifying weak spectral features in observational data, the model data should
be first convolved to the spectral resolution of the telescope and the
appropriate amount of noise be added to the model in order to make an accurate
representation as possible to the observed data.  Only then can the 1-2$\sigma$
spectral features be corrected assessed as statistically significant  features
in the observed data and not an artifact of the noise and spectral resolution.

\section{SUMMARY}

We have conducted a high resolution acetic acid (CH$_{3}$COOH) survey at 3 mm
wavelengths for the first time with CARMA towards two high-mass hot cores,
G19.61-0.23 and G29.96-0.02, and a low-mass embedded protostar system, IRAS
16293-2422. We have detected CH$_3$COOH emission in G19.61-0.23, via the two
CH$_3$COOH rotational transitions, $10_{*,10}-9_{*,9}$ $E$ and $A$. While
G19.61-0.23 consists of several clumps, high resolution mapping reveals that the
CH$_3$COOH emission is extremely compact ($<$2\arcsec) toward component C.
Methyl formate (HCOOCH$_3$)  and ethyl cyanide (CH$_3$CH$_2$CN) were also
detected.  The emission from these two molecules is more extended than the
CH$_{3}$COOH emission but the peaks of all three emission spectra are toward
component C. The CH$_{3}$COOH column density and abundance ratio with respect to
methyl formate (HCOOCH$_{3}$), are 2.0(1.0)$\times$10$^{16}$ cm$^{-2}$ and
2.2(0.1)$\times$10$^{-1}$ respectively, which is higher but comparable to the
other high-mass CH$_3$COOH sources given the current level of detection.
Although acetic acid is not detected toward  G29.96-0.02, we have detected
HCOOCH$_{3}$ and CH$_3$CH$_2$CN.  The HCOOCH$_{3}$ and CH$_3$CH$_2$CN emission
features of G29.96-0.02 are  weaker than G19.61-0.23 by a factor of 3. While we
detect CH$_{3}$COOH in  G19.61-0.23, the non-detection of CH$_{3}$COOH in
G29.96-0.02 may be due to the limited sensitivity.  Finally we have detected two
new transitions of HCOOCH$_{3}$, $9_{2,8}-8_{2,7}$ $E$  and $A$ toward IRAS
16293-2422 and weak emission features at the 2$\sigma$ level of CH$_3$COOH
toward I16293A.  This is the first observation of new spectral  features of
CH$_3$COOH since the reported detection by \citet{cazaux03}. The HCOOCH$_{3}$ 
spectrum toward I16293B shows narrower and more distinguishable line profiles
than the spectrum toward I16293A, which may be suffering from self-absorption.
However, we did not  detect CH$_3$CH$_2$CN in IRAS 16293-2422 at the the
1$\sigma$ detection limit. 

The new CH$_3$COOH detection demonstrates the strong correlation between large O
and N-bearing species and CH$_3$COOH in hot cores. The timescale of hot cores
may also play a role in O/N chemical differentiation and hence CH$_3$COOH
formation. However, we need more CH$_3$COOH detections to confirm these
suggestions. The compactness of the emission from large molecules stresses the need of
interferometers for hot core chemistry observations. Our high resolution survey has provided
critical information about the chemical and physical properties of hot cores,
which can be used to constrain the hot core chemical models. Higher resolution
observations are still required to precisely determine temperature and the
density distribution of large O and N-bearing species for physical-parameter
sensitive chemical models. Through well-improved chemical models, we can better
understand the distinctive hot core chemistry.

\acknowledgments

\begin{figure}
\begin{center}
\includegraphics[angle=270,scale=0.4]{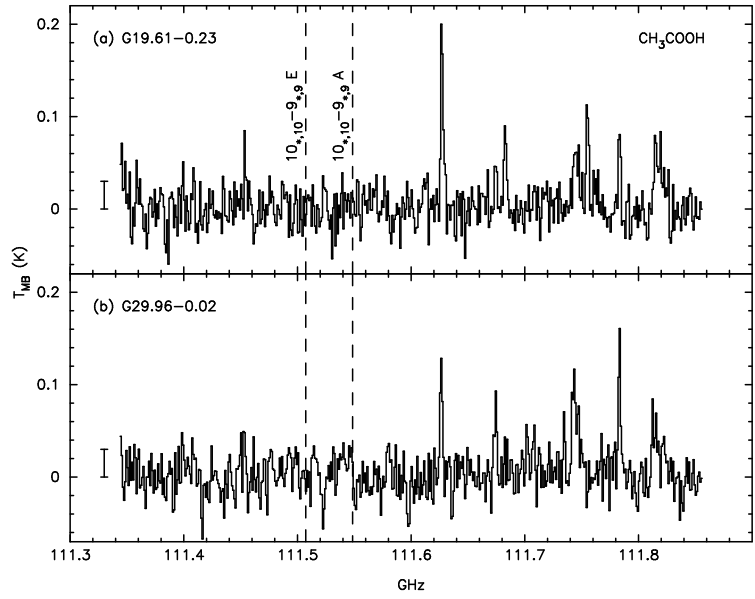}
\caption{IRAM 30-m observations toward G19.61-0.23 and G29.96-0.02
\citep{font07}.  The acetic acid lines are covered in this window at 111.50727
and 111.54853 GHz.}
\label{spec_30m}
\end{center}
\end{figure}

We especially would like to thank Francesco Fontani, Paolo Caselli and Friedrich
Wyrowski for providing the IRAM 30-m data presented in Figure \ref{spec_30m}. 
It is with this type of collaboration that we can truly make advances in
understanding the formation of complex molecules in astronomical environments. 
We also thank an anonymous referee for some  very useful comments that
undoubtedly strengthen the manuscript.  We acknowledge support from the
Laboratory for Astronomical Imaging at the University of Illinois and NSF grant
AST-0540459. Support for CARMA construction was derived from the Gordon and
Betty Moore Foundation, the Kenneth T. and Eileen L. Norris Foundation, the
Associates of the California Institute of Technology, the states of California,
Illinois, and Maryland, and the National Science Foundation. Ongoing CARMA
development and operations are supported by the National Science Foundation
under a cooperative agreement, and by the CARMA partner universities. This work
was also supported in part by the NSF Centers for Chemical Innovation through
award CHE-0847919.

\clearpage
\bibliography{acabib}{}
\clearpage

\begin{deluxetable}{lrrrrcccc}
\tabletypesize{\scriptsize}
\tablecaption{Observed Sources\label{tbl-src}}
\tablewidth{0pt}
\tablehead{ \colhead{Source} & \colhead{$\alpha$} &
\colhead{$\delta$} & \colhead{Flux Calibrator} &
\colhead{Gain Calibrator} & \colhead{Distance} &
\colhead{Mass} & \colhead{$v_{LSR}$} \\
\colhead{} & \colhead{(J2000.0)} &
\colhead{(J2000.0)} & \colhead{} &
\colhead{} & \colhead{(kpc)} &
\colhead{($M_\sun$)} & \colhead{(km s$^{-1}$)}
}
\startdata
 G19.61-0.23\tablenotemark{a}      & 18 27 38.1 & -11 56 39.0   
        & Neptune & 1911-201 & 3.50 & 450 & 40.0 \\
 G29.96-0.02\tablenotemark{b}      & 18 46 04.0  & -02 39 21.5   
        & Neptune & 1751+096 & 6.0 & 1100 & 98.8 \\
 IRAS 16293-2422\tablenotemark{c,d}  & 16 32 22.8 & -24 28 33.0
        & MWC349 & 1625-254 & 0.16 & 5.4 & 3.9  \\
	& & & & 1733-130 & & & \\
\enddata
\tablenotetext{a}{\citet{furu05}}
\tablenotetext{b}{\citet{olmi03}}
\tablenotetext{c}{\citet{schoier02}}
\tablenotetext{d}{\citet{fran01}}
\end{deluxetable}

\begin{deluxetable}{lcclrc}
\tabletypesize{\scriptsize}
\tablecaption{OBSERVED MOLECULES AND TRANSITIONS\label{tbl-mol}}
\tablewidth{0pt}
\tablehead{ \colhead{Species} & \colhead{Rotational Partition Function} 
& \colhead{Transition} & \colhead{Frequency (MHz)} & \colhead{$E_u$ (K)} 
& \colhead{$<S_{j,j}\mu^2>$ (debye$^2$)} }
\startdata
CH$_3$COOH\tablenotemark{a} & $Q_r=14.1T_r^{3/2}$
           & $10_{*,10}-9_{*,9}$ $E$ & 111,507.270(20) &  30.5 & 54.8 \\
         & & $10_{*,10}-9_{*,9}$ $A$ & 111,548.533(20) &  30.5 & 54.8 \\
HCOOCH$_3$\tablenotemark{b} & $Q_r=12.45T_r^{3/2}$
           & $9_{2,8}-8_{2,7}$ $E$ & 107,537.189(25) & 28.8 & 22.8 \\
         & & $9_{2,8}-8_{2,7}$ $A$ & 107,543.746(25) & 28.8 & 22.8 \\
CH$_3$CH$_2$CN\tablenotemark{c} & $Q_r=7.17T_r^{3/2}$
                & $12_{11,2}-11_{11,1}$ & 107,539.854 (3)  & 167.8 & 28.4 \\       
             &  & $12_{11,1}-11_{11,0}$ & 107,539.854 (3)  & 167.8 & 28.4 \\
             &  & $12_{4,9}-11_{4,8}$  & 107,543.918  (3) & 51.4 & 158.1 \\
             &  & $12_{4,8}-11_{4,7}$  & 107,547.593  (3) & 51.4 & 158.1 \\
             &  & $12_{3,10}-11_{3,9}$ & 107,594.040  (3) & 43.6 & 166.8 \\
\enddata
\tablecomments{The 1$\sigma$ uncertainty of the frequencies is in units of kHz. Each of
the CH$_3$COOH lines consists of two a-type and two b-type transitions. This is
represented by an asterisk substituted for the K$_{-}$ quantum numbers.}
\tablenotetext{a}{\citet{ilyu08}}
\tablenotetext{b}{\citet{oest99}}
\tablenotetext{c}{\citet{fuku96,remi07}}
\end{deluxetable}

\begin{deluxetable}{lccccc}
\tabletypesize{\scriptsize}
\tablecaption{DETECTED MOLECULES AND TRANSITIONS TOWARD G19.61-0.23
    \label{tbl-g19}}
\tablewidth{0pt}
\tablehead{ \colhead{Species} & \colhead{Transition} &
\colhead{$\Delta$I (Jy beam$^{-1}$)} & \colhead{$\Delta\upsilon$ (km s$^{-1}$)}& 
\colhead{$\Delta$I (Jy beam$^{-1}$)} & \colhead{$\Delta\upsilon$ (km s$^{-1}$)}}
\startdata
& & \multicolumn{2}{c}{Fitted Data} & \multicolumn{2}{c}{Model Data}\\
\hline
CH$_3$COOH & $10_{*,10}-9_{*,9}$ $E$ & 0.07$\pm$0.01 & 6.4$\pm$0.9 & 0.10 & 6.4\\
           & $10_{*,10}-9_{*,9}$ $A$ & 0.09$\pm$0.02 & 6.4$\pm$0.9 & 0.12 & 6.4 \\
HCOOCH$_3$ & $9_{2,8}-8_{2,7}$ $E$   & 0.20$\pm$0.02 &  6.4$\pm$0.9 & 0.18 & 6.4\\
           & $9_{2,8}-8_{2,7}$ $A$    & 0.20$\pm$0.02 & 6.4$\pm$0.9 & 0.18 & 6.4\\
CH$_3$CH$_2$CN & $12_{11,*}-11_{11,*}$ & 0.18$\pm$0.02 & 12.9$\pm$0.4 & 0.13 & 12.9\\
               & $12_{4,9}-11_{4,8}$  & 0.59$\pm$0.02 & 12.9$\pm$0.4 & 0.66 & 12.9\\
               & $12_{4,8}-11_{4,7}$  & 0.59$\pm$0.02 & 12.9$\pm$0.4 & 0.66 & 12.9\\
               & $12_{3,10}-11_{3,9}$ & 0.71$\pm$0.01 & 12.9$\pm$0.4 & 0.72 & 12.9\\
U107.588 & 107.5880 GHz & 0.16$\pm$0.04 & 5.6$\pm$1.6 & & \\
U107.591 & 107.5910 GHz & 0.27$\pm$0.03 & 7.8$\pm$1.2 & & \\
U107.597 & 107.5970 GHz & 0.31$\pm$0.03 & 12.5$\pm$1.2 & & \\
U107.604 & 107.6040 GHz & 0.23$\pm$0.03 & 11.1$\pm$1.5 & & \\
U111.506 & 111.5055 GHz & 0.06$\pm$0.02 & 5.2$\pm$1.7 & & \\
U111.510 & 111.5095 GHz & 0.07$\pm$0.01 & 17.2$\pm$4.5 & & \\
U111.546 & 111.5460 GHz & 0.07$\pm$0.01 & 8.7$\pm$2.0 & & \\

\enddata
\tablecomments{The HCOOCH$_3$ lines were fit assuming the same linewidth.  The
same line width fitting criteria applies for the CH$_3$CH$_2$CN and CH$_3$COOH
lines. For simplicity, the two CH$_3$CH$_2$CN lines listed in Table 2 are now
represented as one by an asterisk substituted for the K$_{-}$ quantum numbers. 
The other three CH$_3$CH$_2$CN lines remain unchanged.}
\end{deluxetable}

\begin{deluxetable}{lccccc}
\tabletypesize{\scriptsize}
\tablecaption{DETECTED MOLECULES AND TRANSITIONS TOWARD G29.96-0.02
    \label{tbl-g29}}
\tablewidth{0pt}
\tablehead{ \colhead{Species} & \colhead{Transition} &
\colhead{$\Delta$I (Jy beam$^{-1}$)} & \colhead{$\Delta\upsilon$ (km s$^{-1}$)}& 
\colhead{$\Delta$I (Jy beam$^{-1}$)} & \colhead{$\Delta\upsilon$ (km s$^{-1}$)}}
\startdata
& & \multicolumn{2}{c}{Fitted Data} & \multicolumn{2}{c}{Model Data}\\
\hline
CH$_3$COOH & $10_{*,10}-9_{*,9}$ $E$ & $\sim$0.04 &  &  & \\
           & $10_{*,10}-9_{*,9}$ $A$ & $<$0.03    &  &  & \\
HCOOCH$_3$ & $9_{2,8}-8_{2,7}$ $E$   & 0.08$\pm$0.01 &  7.8$\pm$1.1 & 0.09 & 7.8\\
           & $9_{2,8}-8_{2,7}$ $A$    & 0.08$\pm$0.01 & 7.8$\pm$1.1 & 0.09 & 7.8\\
CH$_3$CH$_2$CN & $12_{11,*}-11_{11,*}$ & 0.07$\pm$0.01 & 10.8$\pm$0.4 & 0.03 & 10.8\\
               & $12_{4,9}-11_{4,8}$  & 0.21$\pm$0.01 & 10.8$\pm$0.4 & 0.23 & 10.8\\
               & $12_{4,8}-11_{4,7}$  & 0.21$\pm$0.01 & 10.8$\pm$0.4 & 0.23 & 10.8\\
               & $12_{3,10}-11_{3,9}$ & 0.27$\pm$0.01 & 10.8$\pm$0.4 & 0.26 & 10.8\\
U107.591 & 107.5910 GHz & 0.11$\pm$0.01 & 8.5$\pm$0.9 & & \\
U107.597 & 107.5975 GHz & 0.08$\pm$0.01 & 10.8$\pm$1.2 & & \\
U107.604 & 107.6040 GHz & 0.10$\pm$0.01 & 12.2$\pm$1.3 & & \\
U111.541 & 111.5410 GHz & 0.07$\pm$0.01 & 12.7$\pm$2.0 & & \\

\enddata
\tablecomments{Table comments are the same as in Table 3.}
\end{deluxetable}

\begin{deluxetable}{lccccc}
\tabletypesize{\scriptsize}
\tablecaption{DETECTED MOLECULES AND TRANSITIONS TOWARD IRAS 16293-2422
\label{tbl-i16293}}
\tablewidth{0pt}
\tablehead{\colhead{Species} & \colhead{Transition} &
\colhead{$\Delta$I (Jy beam$^{-1}$)} & \colhead{$\Delta\upsilon$ (km s$^{-1}$)}&
\colhead{$\Delta$I (Jy beam$^{-1}$)} & \colhead{$\Delta\upsilon$ (km s$^{-1}$)}}
\startdata
& & \multicolumn{2}{c}{Fitted Data} & \multicolumn{2}{c}{Model Data}\\
\hline
\multicolumn{6}{c}{IRAS 16293-2422 A} \\
\hline
CH$_3$COOH & $10_{*,10}-9_{*,9}$ $E$ & 0.05$\pm$0.01 & 6.0$\pm$2.0  & 0.03 & 6\\
           & $10_{*,10}-9_{*,9}$ $A$ & 0.05$\pm$0.01 & 6.0$\pm$2.0  & 0.05 & 6\\
HCOOCH$_3$& $9_{2,8}-8_{2,7}$ $E$ & 0.050$\pm$0.009 & 14.6$\pm$1.5 & 0.049 & 14.6\\
          & $9_{2,8}-8_{2,7}$ $A$ & 0.050$\pm$0.008 & 14.6$\pm$1.5 & 0.049 & 14.6\\
U111.508 & 111.5080 GHz & 0.05$\pm$0.01 & 5.0$\pm$2.0 & & \\
\hline
\multicolumn{6}{c}{IRAS 16293-2422 B} \\
\hline
CH$_3$COOH & $10_{*,10}-9_{*,9}$ $E$ & $<$0.01 &  &  & \\
           & $10_{*,10}-9_{*,9}$ $A$ & $<$0.01 &  &  & \\
HCOOCH$_3$& $9_{2,8}-8_{2,7}$ $E$ & 0.063$\pm$0.005 & 5.5$\pm$0.5 & 0.059 & 5.5\\
          & $9_{2,8}-8_{2,7}$ $A$ & 0.063$\pm$0.005 & 5.5$\pm$0.5 & 0.059 & 5.5\\
U111.510 & 111.5095 GHz & 0.06$\pm$0.01 & 8.0$\pm$2.0 & & \\
U111.541 & 111.5410 GHz & 0.06$\pm$0.01 & 5.0$\pm$1.0 & & \\
\enddata
\end{deluxetable}

\begin{deluxetable}{lrrrr}
\tabletypesize{\scriptsize}
\tablecaption{COLUMN DENSITIES OF THE OBSERVED MOLECULES \label{tbl-col}}
\tablewidth{0pt}
\tablehead{ \colhead{Species (cm$^{-2}$)} & \colhead{G19.61-0.23} & 
         \colhead{G29.96-0.02} & \colhead{I16293A} & \colhead{I16293B}}
\startdata
CH$_3$COOH  & (2.0$\pm$1.0)$\times$10$^{16}$ & $<$ 9$\times$10$^{14}$ & $\sim$1.6$\times$10$^{15}$ & $<$6$\times$10$^{14}$ \\
HCOOCH$_3$  & (9.0$\pm$2.0)$\times$10$^{16}$ & (4.0$\pm$1.0)$\times$10$^{16}$ &  (1.5$\pm$0.3)$\times$10$^{16}$  & (6.0$\pm$1.0)$\times$10$^{15}$\\    
CH$_3$CH$_2$CN & (6.0$\pm$3.0)$\times$10$^{16}$ &(1.1$\pm$0.7)$\times$10$^{16}$ & $<$6$\times$10$^{13}$ & $<$3$\times$10$^{14}$ \\
\tablecomments{The column densities are in units of cm$^{-2}$.}
\enddata
\end{deluxetable}

\begin{deluxetable}{lccccc}
\tabletypesize{\scriptsize}
\tablecaption{COMPARISON OF CH$_{3}$COOH SOURCES \label{tbl-aca_cmp}}
\tablewidth{0pt}
\tablehead{ \colhead{Species} & \colhead{Sgr B2(N-LMH)\tablenotemark{a}} &
\colhead{W51e2\tablenotemark{b}} & \colhead{G34.3+0.15\tablenotemark{c}} 
& \colhead{G19.61-0.23\tablenotemark{d}} & \colhead{IRAS 16293 A\tablenotemark{d}}}
\startdata
N$_\mathrm{CH_3COOH}$ (cm$^{-2}$) & $\sim$7.3$\times$10$^{15}$  
          & 1.7$\pm$0.5$\times$10$^{16}$  & (0.77-1.64)$\times$10$^{15}$
          & 2.0$\pm$1.0$\times$10$^{16}$ & $\sim$1.6$\times$10$^{15}$ \\
N$_\mathrm{CH_3COOH}$/N$_\mathrm{HCOOCH_3}$  & (4-7)$\times$10$^{-2}$    &
                          (1-6)$\times$10$^{-2}$ & $\sim$3.3$\times$10$^{-2}$ & 
			  2.2$\pm$0.1$\times$10$^{-1}$ & 
			  1.0$\pm$0.1$\times$10$^{-1}$ \\
\enddata
\tablenotetext{a}{\citet{meh97}; $T_r=$200 K}
\tablenotetext{b}{\citet{remi02}; $T_r=$201 K}
\tablenotetext{c}{\citet{remi03}; $T_r=$70-185 K}
\tablenotetext{d}{this work; $T_r=$161 K for G19 and $T_r=$62 K for I16293A}
\end{deluxetable}

\clearpage

\end{document}